\documentclass[aps,prd,twocolumn,nofootinbib]{revtex4-1}
\usepackage{graphicx}

\newcommand{\tmtexttt}[1]{{\ttfamily{#1}}}

\begin{document}

\title{GENXICC2.1: An Improved Version of GENXICC for Hadronic Production of Doubly Heavy Baryons}

\author{Xian-You Wang}
\email{xywang@cern.ch}
\affiliation{Department of Physics, Chongqing University, Chongqing 401331, P.R. China \\ Institute of High Energy Physics, Chinese Academy of Sciences, Beijing 100049, P.R. China}

\author{Xing-Gang Wu}
\email{wuxg@cqu.edu.cn}
\affiliation{Department of Physics, Chongqing University, Chongqing 401331, P.R. China}

\date{\today}

\begin{abstract}
We present an improved version of GENXICC, which is a generator for hadronic production of the doubly heavy baryons $\Xi_{cc}$, $\Xi_{bc}$ and $\Xi_{bb}$ and has been raised by C.H. Chang, J.X. Wang and X.G. Wu [Comput. Phys. Commun. {\bf 177} (2007) 467; Comput. Phys. Commun. {\bf 181} (2010) 1144]. In comparison with the previous GENXICC versions, we update the program in order to generate the unweighted baryon events more effectively under various simulation environments, whose distributions are now generated according to the probability proportional to the integrand. One Les Houches Event (LHE) common block has been added to produce a standard LHE data file that contains useful information of the doubly heavy baryon and its accompanying partons. Such LHE data can be conveniently imported into PYTHIA to do further hadronization and decay simulation, especially, the color-flow problem can be solved with PYTHIA8.0.
\end{abstract}

\maketitle

\noindent{\bf NEW VERSION PROGRAM SUMMARY} \\

\noindent{\it Title of program} : GENXICC2.1 \\

\noindent{\it Program obtained from} : CPC Program Library \\

\noindent{\it Reference to original program} : GENXICC \\

\noindent{\it Reference in CPC} : Comput. Phys. Commun. {\bf 177}, 467 (2007); Comput. Phys. Commun. {\bf 181}, 1144 (2010) \\

\noindent{\it Does the new version supersede the old program}: No \\

\noindent{\it Computer} : Any LINUX based on PC with FORTRAN 77 or FORTRAN 90 and GNU C compiler as well \\

\noindent{\it Operating systems} : LINUX \\

\noindent{\it Programming language used} : FORTRAN 77/90\\

\noindent{\it Memory required to execute with typical data} : About 2.0 MB \\

\noindent{\it No. of bytes in distributed program} : About 2 MB, including PYTHIA6.4 \\

\noindent{\it Distribution format} : .tar.gz \\

\noindent{\it Nature of physical problem} : Hadronic production of doubly heavy baryons $\Xi_{cc}$, $\Xi_{bc}$ and $\Xi_{bb}$. \\

\noindent{\it Method of solution} : The upgraded version with proper interface to PYTHIA can generate full production and decay events, either weighted or unweighted, conveniently and effectively. Especially, the unweighted events are generated by using an improved hit-and-miss approach. \\

\noindent{\it Reasons for new version} : Responding to the feedback from users of CMS and LHCb groups at the large hadronic collider, and basing on the recent improvements of PYTHIA on the color-flow problem, we improve the efficiency for generating the unweighted events, and also improve the color-flow part for further hadronization. Especially, an interface has been added to import the output production events into a suitable form for PYTHIA8.0 simulation, in which the color-flow during the simulation can be correctly set. \\

\noindent{\it Typical running time} : It depends on which option is chosen to match PYTHIA when generating the full events and also on which mechanism is chosen to generate the events. Typically, for the dominant gluon-gluon fusion mechanism to generate the mixed events via the intermediate diquarks in $(cc)[^3S_1]_{\bar{3}}$ and $(cc)[^1S_0]_6$ states, setting IDWTUP=3 and \tmtexttt{unwght}=.true., it takes 30 minutes to generate $10^5$ unweighted events on a 2.27GHz Intel Xeon E5520 processor machine; setting IDWTUP=3 and \tmtexttt{unwght}=.false. or IDWTUP=1 and IGENERATE=0, it only needs 2 minutes to generate the $10^5$ baryon events (the fastest way, for theoretical purpose only). As a comparison, for previous GENXICC versions, if setting IDWTUP=1 and IGENERATE=1, it takes about 22 hours to generate 1000 unweighted events. \\


\noindent{\it Summary of the changes (improvements)} : 1) The scheme for generating unweighted events has been improved; 2) One Les Houches Event (LHE) common block has been added to record the standard LHE data in order to be the correct input for PYTHIA8.0 for later simulation; 3) We present the code for connecting GENXICC to PYTHIA8.0, where three color-flows have to be correctly set for later simulation. More specifically, we present the changes together with their detailed explanations in the following :

\begin{itemize}
\item {\bf Unweighted events generation}. For theoretical studies, e.g. to derive the total baryon production cross-section or various differential distributions, one can directly use the fastest way, e.g. setting the PYTHIA parameter IDWTUP=3 and \tmtexttt{unwght}=.false. or setting IDWTUP$=1$ and IGENERATE=0 (in these cases, \tmtexttt{xmaxup} should be set as 0), to generate the baryon events~\cite{pythia6}. By using GENXICC~\cite{genxicc1,genxicc2} in this way, some interesting properties for hadronic production of $\Xi_{cc}$, $\Xi_{bc}$ and $\Xi_{bb}$ have been found in the literature, cf. Refs.\cite{usegen1,usegen2,usegen3}. While, for the events simulation in detector conditions, it is necessary to get the unweighted events. In previous GENXICC versions, the unweighted events are generated by setting IDWGTUP $=1$ and IGENERATE$=1$; i.e., the events are generated according to PYTHIA's inner mechanism, the so-called hit-and-miss approach (von Neumann algorithm), to reject those unsatisfied events and output the allowed events. But, as is well-known, the original hit-and-miss approach is really time-consuming. Some alterations must be made to improve its efficiency.

    As an intermediate step, in BCVEGPY2.1a~\cite{bcvegpy21a} we have suggested a practical trick to increase the efficiency of generating unweighted events (BCVEGPY is a generator for hadronic production $B_c$ meson~\cite{bcvegpy}). In this trick, other than choosing the maximum differential cross-section as a reference weight in the hit-and-miss approach, we directly select an effective differential cross-section, which is smaller than the maximum one, as the reference weight~\cite{bcvegpy21a}. This treatment can largely improve the generation efficiency without affecting the total cross-section of the process. However, in using this trick to generate unweighted events such as for CMS detector simulation, one will incidentally find a false peak in the $B_c$-$p_t$ distributions. This is caused by the fact that sometimes the same event will be stored with (false) large number of times in the hit-and-miss process. Then, we are facing a dilemma: such a false peak can be avoided by rising the effective reference weight to a value approaching the maximum weight, but, inversely, a larger reference weight will surely lead to a much longer running-time.

    One observes that by using the VEGAS algorithm~\cite{vegas}, the SPRING-BASES program~\cite{basespring} performs the integration in using the BASES subroutines and generates events with a probability proportional to the integrand in using the SPRING subroutines. After each iteration of VEGAS running, the integration result and the maximum value of the function will be stored in a file for each cell of the adaptive mesh. In the generation stage, a cell is chosen with a probability proportional to the corresponding integral, and then a point in the cell is generated using the hit-and-miss approach. This method is highly efficient, but it has the disadvantage that the required amount of storage space grows exponentially with the integration dimension.

    Next, in POWHEG program~\cite{powheg} the authors have developed a new method MINT~\cite{mint} to replace the SPRING-BASES package. This MINT package also use the VEGAS algorithm to perform the integration. What's the difference is that it does not store the value of the integral but stores the upper bound value for each cell. The multidimensional stepwise function that equals to the upper bound of the function to be integrated in each cell is in fact an upper bound for the whole function, which is the wanted upper bound for BCVEGPY2.1a or the PYTHIA. So, the program is to find the upper bound grid for those cells. And next, by using again the hit-and-miss technique in each cell, one can generate the points according to the original distribution.

    Basing on these methods, as a further improvement, we present an ultimate solution to generate unweighted events in the present new GENXICC version. We adopt the MINT algorithm but with certain alterations to do the simulation. For the purpose, we change the VEGAS subroutine as follows. Three new variables have been added in the original VEGAS subroutine, where \tmtexttt{xint} is the integral value for the integrand \tmtexttt{fxn} after a \tmtexttt{ndim}-dimensional integration, the \tmtexttt{xmax} array records the upper bounding envelope of the integrand in all cells, \tmtexttt{imode} is a flag :
    \begin{widetext}
    \centering
    \tmtexttt{vegas(fxn,ndim,ncall,itmx,nprn,xint,xmax,imode)}
    \end{widetext}
    \begin{itemize}
      \item When called with \tmtexttt{imode=0}, \tmtexttt{vegas} performs the integration over the integrand \tmtexttt{fxn}, and stores the answer in a common block parameter \tmtexttt{vegsec}.
      \item \tmtexttt{xmax} stands for a (\tmtexttt{nvegbin},\tmtexttt{ndim}) dimensional array, where \tmtexttt{nvegbin} denotes the bin number 
          for each coordinate, \tmtexttt{ndim} stand for the integration dimension. When called with \tmtexttt{imode=1}, \tmtexttt{vegas} will first initiate all the elements of \tmtexttt{xmax} to be ${\rm xint}^{1/{\rm ndim}}$, where \tmtexttt{xint} equals to the value of \tmtexttt{vegsec} that has been derived from a previous VEGAS running with \tmtexttt{imode=0}. During the following sampling iteration, when the calculated integral value is larger than the initial \tmtexttt{xmax}(\tmtexttt{nvegbin},\tmtexttt{ndim}) value in a specific cell, then the value of \tmtexttt{xmax}(\tmtexttt{nvegbin},\tmtexttt{ndim}) for this cell will be increased by a fixed factor $f=1+1/10\,{\rm ndim}$. After a sufficiently large number of calls, the values of \tmtexttt{xmax}(\tmtexttt{nvegbin},\tmtexttt{ndim}) will be stabilized for all cells. Such a final \tmtexttt{xmax} array will be stored in the same grid file as that of the importance sampling function in order to do the final simulation.
    \end{itemize}

     Comparing to the previous GENXICC versions, in doing the initialization (subroutine \tmtexttt{evntinit}), we will call \tmtexttt{vegas} twice with \tmtexttt{imode=0} and \tmtexttt{imode=1} accordingly to generate the upper bound grid \tmtexttt{xmax} and also a more precise importance sampling function. Practically, the user can directly use the existed grid file derived by previous VEGAS running to generate events by setting \tmtexttt{methodevnt=2} or \tmtexttt{methodevnt=3} without running VEGAS again, which is the same as the older GENXICC versions.

     Once the \tmtexttt{xmax} array has been set up in previous steps, one can call the subroutine \tmtexttt{gen} to generate events. For the purpose, three options for calling \tmtexttt{gen} subroutine are programmed : \\
     \tmtexttt{\\
      jmode=0\\
      call gen(fxn,ndim,xmax,jmode)\\
      jmode=1\\
      do j=1,10000\\
      call gen(fxn,ndim,xmax,jmode)\\
      ...\\
      enddo\\
      jmode=3\\
      call gen(fxn,ndim,xmax,jmode)\\}\\
    where \tmtexttt{jmode=0} is to initializes a step-wise function \tmtexttt{xmmm} which descripted in\cite{mint}. And \tmtexttt{jmode=3} is to print out the generation statistics.

    The calling for the \tmtexttt{gen} subroutine with \tmtexttt{jmode=1} has been implemented into the UPEVNT subroutine to generate events according to the probability proportional to the integrand. Each event produced needs several times of iteration with three steps procedure as follows:
     \begin{enumerate}
     \item Calculate upper bounding function by generating a set of step-wise functions, each of them is associated with a specific coordinate (dimension).
     \item Call the \tmtexttt{phase$\_$gen} subroutine to generate a random phase-space point and calculate the integral.
      \item Judge whether such point be kept or not by using the hit-and-miss approach with the help of the upper bounding function.
     \end{enumerate}

    In VEGAS the integral together with its numerical error are related to the sampling numbers \tmtexttt{ncall} and the iteration times \tmtexttt{itmx}. So, to generate full events, we suggest the user to do a test running first in order to find an effective and time-saving parameters for VEGAS. Furthermore, to validate the program, we use the same default parameters as the input for the program to generate mixed events via the intermediate diquark in $(cc)[^3S_1]_{\bar{3}}$ and $(cc)[^1S_0]_6$ states, and the same for other two doubly heavy baryons $\Xi_{bc}$ and $\Xi_{bb}$.

    \begin{figure}[tb]
    \centering
    \includegraphics[width=0.45\textwidth]{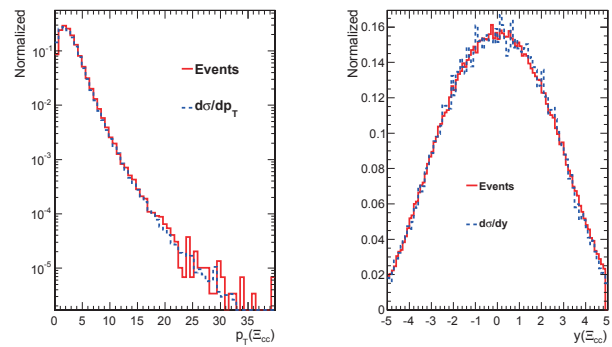}
    \caption{Comparison of the normalized $\Xi_{cc}$ transverse momentum ($P_T$) and rapidity ($y$) distributions derived by IDWTUP=3 (events) and IDWTUP=1 (differential cross-sections), which are represented by solid line and dotted line respectively. }
    \label{fig1}
    \end{figure}

    As a cross-check, we derive the unweighted $\Xi_{cc}$ event distributions by setting IDWTUP=3 and \tmtexttt{unwght}=.true., and the weighted $\Xi_{cc}$ differential distributions by setting IDWTUP=1 and IGENERATE=0, respectively, which are shown in FIG.\ref{fig1}. The two distributions after proper normalization agree well with each other. This demonstrates that our present scheme for unweighted events is correct.

\item {\bf Color-flow problem.} Within the framework of non-relativistic QCD (NRQCD)~\cite{nrqcd}, the production of $\Xi_{QQ^{\prime}}$ baryon can be factorized into two steps: The first step is to produce two free heavy-quark pairs $Q\bar{Q}$ and $Q^{\prime}\bar{Q}^{\prime}$, which is perturbatively calculable. The second step is to make the two heavy quarks $Q$ and $Q^{\prime}$ into a bounding diquark $(QQ^{\prime})$ in $[^3S_1]$ (or $[^1S_0]$) spin state and in $\mathbf{\bar{3}}$ (or $\mathbf{6}$) color state accordingly; then it will be hadronized into $\Xi_{QQ^{\prime}}$ baryon by grabbing a light quark $u$ or $d$ or $s$ (plus suitable number of gluons), whose probability is described by non-perturbative NRQCD matrix element. More explicitly, the intermediate diquarks in $\Xi_{cc}$ and $\Xi_{bb}$ have two spin-and-color configurations $[^3S_1]_{\bf\bar{3}}$ and $[^1S_0]_{\bf 6}$; while for the intermediate diquark $(bc)$ in $\Xi_{bc}$, there are four spin-and-color configurations $\Xi_{bc}[^3S_1]_{\bf\bar{3}}$, $\Xi_{bc}[^3S_1]_{\bf 6}$, $\Xi_{bc}[^1S_0]_{\bf\bar{3}}$, and $\Xi_{bc}[^1S_0]_{\bf 6}$.

    Since a baryon is constructed by three valance quarks, under the standard color-flow decomposition, there must be three different color-flow lines being ended at a baryon~\cite{color,genxicc2}. It is different from the case of meson, where the color-flow lines of the quark and anti-quark inside a meson are continued. The previous PYTHIA6.4 can only generate full events with two or less independent color-flow lines, thus in GENXICC2.0, we adopt a `cheating method' to generate the events. That is, by using the fact that $3 \bigotimes 3=6 \bigoplus {\bf \bar{3}}$ and $3 \bigotimes \bar{3}=8 \bigoplus {\bf 1}$ in general QCD SU(3) color space~\cite{genxicc2} :
    \begin{itemize}
     \item We combine any two of the color-flow lines ended with two quarks into one anti-color-flow line ended with one anti-quark with a color $\bar{3}$ that is different from the two quarks (the third color in respect to those of the two quarks);
     \item Secondly, such anti-color-flow line obtained by the combination may be continued (connected) to the remaining quark's color-flow line in the baryon;
     \item Finally, as a consequence, the color-flow lines ended at a baryon become `joined without ends' at all, which is the requirement of the color-singlet bound state.
    \end{itemize}

    However we should point out that due to approximation and simplification with `cheating method', the obtained information about the 'tiny jets', corresponding to the soft anti-quark and soft gluon(s) produced in fragmentation of doubly heavy diquark, may not be very reliable. When the experiment analyzer uses the generator to simulate the baryon decay and other parton hardronization, they are still facing the color-flow rearrangement error; sometimes, PYTHIA will present an error message to show that the color-flow rearrangement is wrong during the parton's evolution process, and then it will stop running.

    To generate full events of the doubly heavy baryons smoothly, the best way is to improve PYTHIA with proper treatment on the color-flow lines ended at the baryon. Fortunately, such an improvement has been done in its newest version PYTHIA8.0. Based on the suggestion from Peter Skands, we find that the further event simulation can be implemented into PYTHIA8.0 correctly even with the previous generated Les Houches Event (LHE) files~\cite{pythia8}. As has been described in Ref.\cite{pythia8BSM}, the read-in of external generator's LHE files generated by PYTHIA6.4 is simply technically less sophisticated and less able to deal with junctions, even though the physics implementation of junction fragmentation is in principle the same. As an solution, PYTHIA8.0 improves the treatment on these LHE files.

    We adopt the same trick as that of BCVEGPY2.1a~\cite{bcvegpy21a} to generate and record the data; i.e. two subroutines have been introduced in the file pythia\/lheinit.F. One of the subroutine XICC\_PYUPIN is used to fill the HEPRUP common block with information on the incoming beams and the allowed processes, and optionally stores that information on file. Another subroutine XICC\_WRITE\_LHE is used to store event information in the HEPEUP common block. And these two subroutines are called by the main program xicc.F to generate the LHE file that records the momentum and color information for the events~\cite{lhe}.

    More specifically, in the main program, the subroutine UPEVNT will be called for generating the baryon events, which is used to call the program to generate the baryon with a probability proportional to the importance sampling function. Here, one can also use the PYTHIA subroutine PYEVNT for the purpose, but one should at the same time switch off the hadronization, the initial and final state parton shower and so on, in order to avoid the color-flow rearrangement error. Those generated baryon information together with the information of the accompanying partons will be stored in the Les Houches common block and will be export to a LHE file ``GENXICC.lhe''. And then, such LHE file can be used when necessary by PYTHIA8.0 to do the following simulations. Here, to successfully simulate the baryon's production and decay, the user need to install the PYTHIA8\cite{pythia8} following the instruction of its official web-site :
    \begin{center}
    \quad\quad http://home.thep.lu.se/~torbjorn/Pythia.html.
    \end{center}
    For using our generator, the user can use the following command to compile the configuration file,
    \begin{widetext}
     \tmtexttt{
      g++ -O2 -ansi -pedantic -W -Wall -Wshadow -I\$(PYTHIA8)/include genxicc.cc -o bin/genxicc.exe \
      -L\$(PYTHIA8)/lib/archive -lpythia8 -llhapdfdummy
     }
    \end{widetext}
    where the \tmtexttt{\$(PYTHIA8)} stands for the PYTHIA8.0 installation directory. 
    
    For convenience, we put an example configuration file in the package for generating the full the baryon production and decay events in PYTHIA8.0, which is placed in the main folder of the program and is named as ``genxicc.cc''.
\end{itemize}

\vspace{1cm}

{\bf Acknowledgments}: We thank Prof.C.H. Chang for helpful discussions on GENXICC. We thank the help of S.Mrenna and P. Skands for the use of PYTHIA8.0, and we thank Z. Li and B. Zhu for helpful discussions on improving the generation efficiency. This work was supported in part by the Fundamental Research Funds for the Central Universities under Grant No. CDJXS1102209 and the Natural Science Foundation of China under Grant No.11075225 and No.11275280, and by the Program for New Century Excellent Talents in University under Grant NO.NCET-10-0882. \\

\end{document}